\documentclass[namedreferences,hyperref,optionalrh,solaromanenum]{spr-sola}

\usepackage{graphicx}                    % For eps figures, newer & more powerfull
\usepackage{color}                       % For color text: \color command
\usepackage{breakurl}                         % For breaking URLs easily trough lines in DVI mode
\newcommand{\aap}{{\it Astron. Astrophys.}}

\newcommand{\apj}{{\it Astrophys. J.}}

\newcommand{\mnras}{{\it Mon. Not. R. Astron. Soc.}}

\newcommand{\pasp}{{\it Publ. Astron. Soc. Pac.}}
\newcommand{\pasj}{{\it Publ. Astron. Soc. Japan}}

\newcommand{\solphys}{{\it Sol. Phys.}}
 
\newcommand{\ssr}{{\it Space Sci. Rev.}} 
\chardef\us=`\_

\newcommand{\araa}{{\it Ann. Rev. Astron. Astrophys.}}

\begin{document}

\begin{frontmatter}

\title{Evaluation of Sunspot Areas Derived by Automated Sunspot-Detection Methods}

%%%%%%%%%%%%%%%%%%%%%%%%%%%%%%%%%%%%%%%%%%%%%%%%%%%
%% Authors Names
%
% \author[addressref={},corref,email={}]{\inits{}\fnm{}\snm{}\orcid{}}
\author[addressref={aff1},corref,email={yoichiro.hanaoka@nao.ac.jp}]{\inits{Y.H.}\fnm{Yoichiro}~\snm{Hanaoka}\orcid{0000-0003-3964-1481}}

%%%%%%%%%%%%%%%%%%%%%%%%%%%%%%%%%%%%%%%%%%%%%%%%%%%
%% Runningheads
%
%\runningauthor{}
%\runningtitle{}
\runningauthor{Y. Hanaoka}
\runningtitle{Evaluation of Sunspot Areas}

%%%%%%%%%%%%%%%%%%%%%%%%%%%%%%%%%%%%%%%%%%%%%%%%%%%
%% Affilations 
%% id shold be the same with \author addressref value.
%\address[id={}]{}
\address[id=aff1]{National Astronomical Observatory of Japan, 2-21-1 Osawa, Mitaka, Tokyo, 181-8588, Japan}

%%%%%%%%%%%%%%%%%%%%%%%%%%%%%%%%%%%%%%%%%%%%%%%%%%%
%%% Abstract 
\begin{abstract}
Sunspot-area measurements using digital images captured by two telescopes at the Mitaka campus of the National Astronomical Observatory of Japan are conducted using automated sunspot detection. A comparison between sunspot areas derived from Mitaka and those from the reference data by \citeauthor{2020A&A...640A..78M} ({\it Astron. Astrophys.} {\bf 640,} A78, \citeyear{2020A&A...640A..78M}), who compiled a cross-calibrated daily sunspot-area catalog, revealed that the correlation coefficients between them are high (0.96--0.97), whereas the areas in the Mitaka data are 70 \%--83 \% of those of Mandal et al. The correlation is limited by the differences in observation times and detection capabilities of spots near the limb, with discrepancies in areas arising from different definitions of spot outlines. Given the high correlation and the ease of calibrating area discrepancies with a correction factor, automated sunspot detection appears promising for future sunspot-area measurements. Furthermore, we addressed the measurements of brightness deficit caused by sunspots.
\end{abstract}

%%%%%%%%%%%%%%%%%%%%%%%%%%%%%%%%%%%%%%%%%%%%%%%%%%%
%% Keywords
%

\end{frontmatter}
%-------------------------------------------------

%%%%%%%%%%%%%%%%%%%%%%%%%%%%%%%%%%%%%%%%%%%%%%%%%%%
%% Sections
%
% \section{}%\label{s:?} 
\section{Introduction} \label{sec:intro}

Long-term records of solar phenomena are key to understanding variations in solar activity. Sunspot records obtained with telescopes are available from the early 17th century, and the daily ``international sunspot number'' has been maintained from 1818 based on sunspot counts \citep{2014SSRv..186...35C}. Meanwhile, the daily total sunspot area, the systematic measurement of which was started in 1874 by the Royal Greenwich Observatory \citep[RGO;][]{2013SoPh..288..117W} and has been continued by several observatories to the present, is another long-term indicator of solar activity. Additionally, efforts have been invested to combine monthly area estimations from the pre-RGO era \citep{1870RSPT..160..389D} with later area data \citep{2004SoPh..221..179V, 2016SoPh..291.2931C}. Although the sunspot number has a shortcoming in that an isolated small spot significantly contributes to the result, such a spot has only a small effect on the total sunspot area. Therefore, sunspot area is expected to be a more objective solar activity index than sunspot number. 

Since the period from 1874 to the present is covered by several sunspot-area records obtained by different observatories, including RGO, efforts have been made to composite these records to obtain a single sunspot-area time series \citep{1997SoPh..173..427F, 2001MNRAS.323..223B, 2013MNRAS.434.1713B, 2009JGRA..114.7104B, 2020A&A...640A..78M}. \citet{2020A&A...640A..78M} composited and cross-calibrated records obtained from nine observatories until 2019. Their catalog is the most complete at this point. In addition, the uninterrupted continuation of sunspot-area measurements is important to further extend the sunspot-area time series. However, currently available (as of 2024) sunspot-area data that are updated daily are only those from the Kislovodsk Observing Station of the Pulkovo Observatory, Russia \citep{2007SoSyR..41...81N} and those from the Solar Optical Observing Network \citep[SOON;][]{2018SoPh..293..138G} operated by the United States Air Force. To correct systematic differences among the results from various observatories and reduce errors in individual sunspot-area measurements, more observatories should provide sunspot-area data; under such circumstances, daily data may be lacking. 

The measurement of sunspot areas was terminated at various observatories possibly because manual area measurements require considerable time and effort. The same limitations are encountered in partly automated area measurements, which require significant human intervention. Conversely, if the process to derive sunspot areas can be performed with an automated method that requires less manpower, then the continuation of the sunspot-area measurements in the future will be possible. Initially, automated methods were developed to measure sunspot areas in photographic images \citep[e.g.,][]{1998SoPh..180..109G, 2014SoPh..289.1403T}, but white-light images taken with digital instruments in recent observations are more suitable for automated processing \citep{2016SoPh..291.3081B}. Therefore, automated methods targeting processing of images taken with the Michelson Doppler Imager of the Solar and Heliospheric Observatory \citep{1995SoPh..162..129S} and the Helioseismic and Magnetic Imager of the Solar Dynamics Observatory \citep{2012SoPh..275..207S} have been developed \citep{2006ESASP.617E.130G, 2014SoPh..289.1403T, 2018PASP..130j4503Y, 2024SoPh..299...10B}. 

The National Astronomical Observatory of Japan (NAOJ) started sunspot detection with an automated method in 1998 using digitally acquired solar images with the ``New Sunspot Telescope'' placed at the Mitaka campus of NAOJ \citep{1998RNAOJ...4....1I, 2002AdSpR..29.1565S}. This is a replacement for the sunspot counts in hand-drawn sketches. In 2012, a new imaging system including a green-continuum imager \citep{2020JSWSC..10...41H} was installed into the ``Solar Flare Telescope'' \citep{1995PASJ...47...81S} also located at Mitaka. The data are processed using an improved automated sunspot detection method \citep{2022SoPh..297..158H}. Both methods provide not only the number of sunspots, but also their areas, and they can be operated automatically with less effort. 

This study aims to verify the validity of the areas derived using these automated methods. To this end, we first compared them with the areas in the catalog by \citet{2020A&A...640A..78M} and evaluated the correlation and systematic differences between them. Next, we investigated the factors that limit the correlation and cause systematic differences. For this purpose, we employed additional area data derived from the Debrecen Heliophysical Observatory (DHO) of Hungary using image data taken at Mitaka. These data enabled us to analyze areas derived from the same data using different types of area measurement methods. 

The automated detection of sunspots using digital images also provides direct measurements of brightness deficits due to sunspots. The brightness deficit changes the total solar irradiance, which is presumed to affect Earth's climate \citep[e.g.,][]{2013ARA&A..51..311S}. For the period before irradiance directly measured from space became available, white-light solar images were considered usable for estimating the brightness deficit. However, measuring the brightness distribution in hand-drawn sketches or photographic images is challenging, although sunspot areas have been derived based on them. Therefore, the ``photometric sunspot index'' \citep[PSI;][]{1982SoPh...76..211H}, which is calculated based on sunspot areas, was proposed to represent the brightness deficit. The brightness deficit derived from digital images is useful for verifying the effectiveness of the PSI. Therefore, we present a comparison between the brightness deficit derived using the automated methods and the PSI compiled by \citet{2020A&A...640A..78M}.

In the following sections, we describe the data used in this study in Section \ref{sec:data}, and the results of the comparisons between various datasets are presented in Section \ref{sec:results}. Section \ref{sec:summary} provides the summary and discussion. The measured brightness deficit owing to sunspots is presented in Appendix \ref{sec:appendix}.

\section{Data} \label{sec:data}

In this section, we describe the datasets used to evaluate sunspot areas derived using the automated sunspot detection methods. In addition to the Mitaka datasets, we used the datasets of \citet{2020A&A...640A..78M} and DHO. Each dataset contains daily records of the total area of all sunspots on the disk. In this paper, the phrase ``sunspot area'' means the area inside a penumbra (including umbra), as commonly used in various studies. Whereas some of sunspot-area catalogs present both the total areas inside penumbrae and the areas of umbrae alone, we deal with only the total areas in this study. Furthermore, two types of areas exist: the projected area measured in a solar image (i.e., the apparent area that suffers the foreshortening effect) and the area corrected for foreshortening. The areas are expressed in millionth of the apparent solar disk area for the projected area, or that of the surface area of a hemisphere for the corrected area ($\mu$Hem). All datasets used here present both projected and corrected areas. These are explained below along with their appellations.

NEWSUN: Results from automated sunspot detection using green-continuum images taken with the New Sunspot Telescope at the Mitaka campus of NAOJ, which is a 10-cm refractor and has a 2K$\times$2K CCD camera \citep{1998RNAOJ...4....1I, 2002AdSpR..29.1565S}. The number and area of sunspots are determined from one image taken per day. Automated detection is based on thresholding; that is, regions with brightness lower than an appropriately defined threshold are distinguished as sunspots. Although it has been working since 1998, the quality of the data collected until 2001 was not very good. Therefore, we used data from 2002 and later for analysis. Some problems are known in the results of spot detection; the detected spots often include false spots (they are removed manually), and small true spots near the limb are often missed.

SFTT4: Results from another method of automated sunspot detection using images taken by the Solar Flare Telescope at the Mitaka campus of NAOJ \citep{1995PASJ...47...81S}, where a 12.5-cm refractor (T4) with a 2K$\times$2K CMOS camera was installed as one of the telescope suites for imaging observations at various wavelengths, including the green continuum \citep{2020JSWSC..10...41H}. Several sets of images for the sunspot detection are taken per day, and the daily number and area of sunspots are determined using one of them. The sunspot detection method is also based on thresholding, but it is an improved method in which the problems found for the New Sunspot Telescope were fixed \citep{2022SoPh..297..158H}. To increase the reliability of sunspot detection, five images, which are taken within a short time but still suffer different blurring due to seeing, are used for daily sunspot-detection. Data from 2012 and later are available.

MANDAL: A composited and cross-calibrated sunspot-area catalog covering 1874--2019 compiled by \citet{2020A&A...640A..78M} based on nine individual sunspot-area records. This catalog can be considered a reference; therefore, we compared the data taken at Mitaka with this catalog and evaluated their quality.

MTKDPD: Area data measured by DHO using images taken with the New Sunspot Telescope at Mitaka. DHO had measured sunspot areas since 1974 \citep{2016SoPh..291.3081B, 2017MNRAS.465.1259G}, and the data were published as the ``DPD'' catalog. Sunspot areas were usually derived from their own observations, but to fill the gap, data from other observatories, including Mitaka, were also used (MTKDPD stands for Mitaka data in the DPD catalog). Therefore, the comparison between the areas derived by different methods from identical images became possible using the MTKDPD and NEWSUN data.

The NEWSUN and SFTT4 data are available at \burl{https://solarwww.mtk.nao.ac.jp/en/db\_sunspot.html}. The MTKDPD data were obtained from \burl{http://fenyi.solarobs.epss.hun-ren.hu/en/databases/DPD}.

The datasets and their observation times and area-derivation methods are listed in Table \ref{tbl:tbl1}. These are daily data, and all datasets, except MANDAL, are based on observations carried out during the same time-period in the day (daytime in Japan). The original data used in MANDAL for the period when NEWSUN and SFTT4 data were available were mostly obtained from the records of Kislovodsk and the DPD catalog of DHO. In MANDAL, the adopted DPD data include some of MTKDPD (data obtained during the daytime in Japan); however, such data are exceptional. Therefore, the observation times of MANDAL, mostly during daytime in Europe, are substantially different from those of the other datasets.

The automated sunspot detection methods used for SFTT4 and NEWSUN are based on thresholding. On the other hand, traditionally, the outline of sunspots was defined by human eyes as done by the RGO, and the method adopted by DHO emulated the visual definition of sunspots \citep{1998SoPh..180..109G}. The sunspot areas determined by RGO, which were adopted in MANDAL as the reference, were close to those of the Kislovodsk records and the DPD catalog \citep{2007SoSyR..41...81N, 2018SoPh..293..142B, 2020A&A...640A..78M}. Therefore, the areas in MANDAL and MTKDPD were considered to be derived using similar traditional methods, which provide the consistent areas irrespective of the observation instrument.

\begin{table}
\caption{Datasets used in this paper and their properties.}
\label{tbl:tbl1}
\begin{tabular}{lcccc}
\hline
Dataset & Observation & Analyzed & Observation Time & Area Derivation Method \\
 & Period & Period & & \\
\hline
SFTT4 & 2012-- & 2012--2019 & Daytime in Japan & Automated Thresholding \\
NEWSUN & 1998-- & 2002--2019 & Daytime in Japan & Automated Thresholding \\
MTKDPD\tabnote{Using images taken with the New Sunspot Telescope.} & & 2012--2015 & Daytime in Japan & Traditional \\
MANDAL & 1874--2019 & 2002--2019 & Daytime in Europe & Traditional \\
\hline
\end{tabular}
\end{table}

\section{Results} \label{sec:results}

\subsection{Comparison between MANDAL and NEWSUN/SFTT4}\label{sec:compmn} 

\begin{figure} 
\centerline{\includegraphics[width=0.5\textwidth]{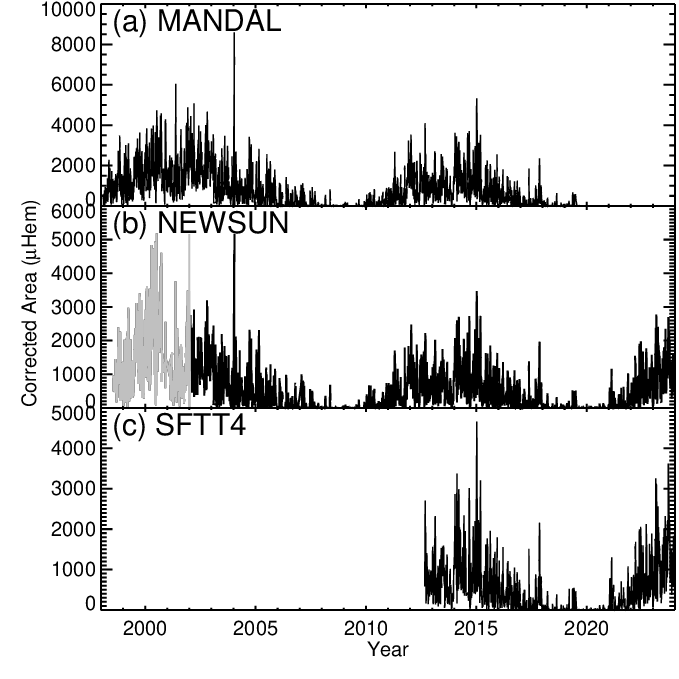}}
\caption{Daily total sunspot areas (foreshortening-effect corrected) of (a) MANDAL, (b) NEWSUN, and (c) SFTT4.  The unit of area is $10^{-6}$ of the area of the hemisphere ($\mu$Hem). For NEWSUN, the data until 2001 (shown in gray in panel (b)) were not used in this study because of their insufficient quality.}\label{fig:fig1}
\end{figure}

\begin{figure} 
\centerline{\includegraphics[width=1\textwidth]{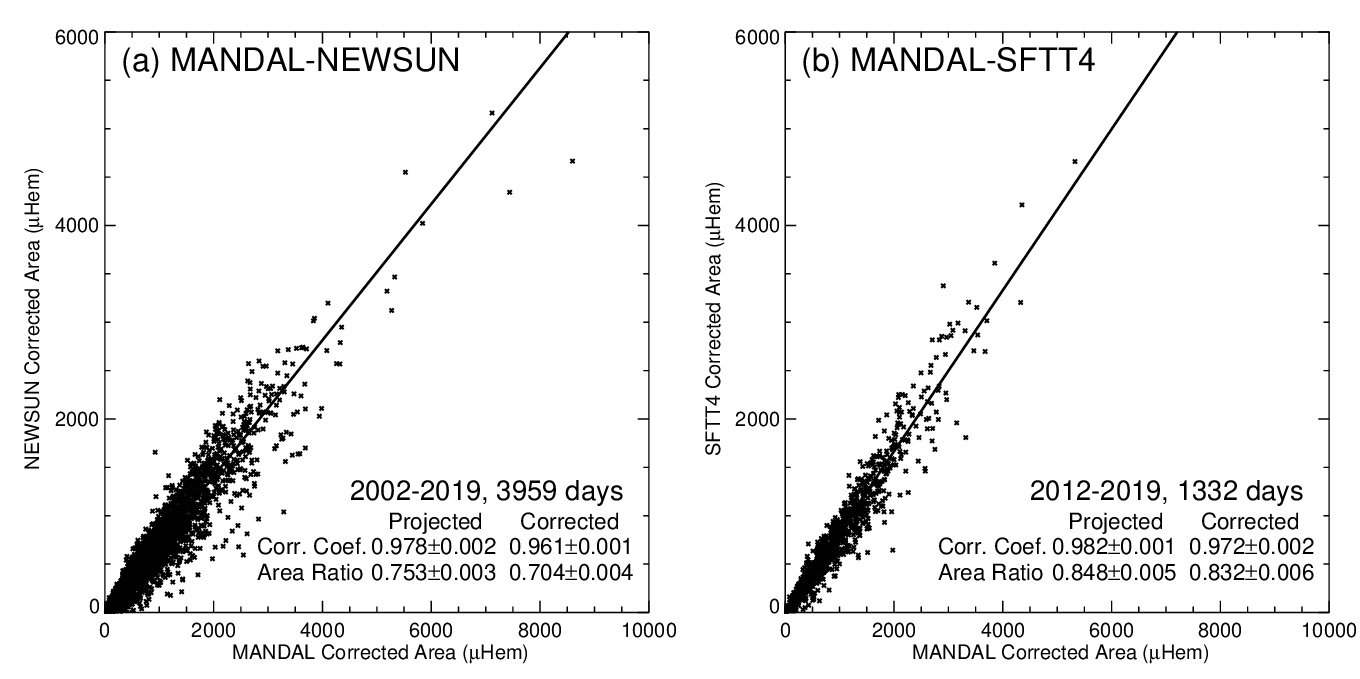}}
\caption{Comparisons of corrected sunspot areas between (a) MANDAL and NEWSUN and (b) MANDAL and SFTT4. Daily data are shown by points, and the regression results are shown by lines. The correlation coefficients and the area ratios (the slopes of the regression lines) for both the projected and corrected areas were annotated.}\label{fig:fig2}
\end{figure}

First, we compare MANDAL and NEWSUN/SFTT4. Figure \ref{fig:fig1} shows the variations in the daily total corrected areas for MANDAL, NEWSUN, and SFTT4. Scatter diagrams of the corrected areas on the common observation days of MANDAL and NEWSUN (2002--2019) and MANDAL and SFTT4 (2012--2019) are presented in Figure \ref{fig:fig2}. The regression results are presented as straight lines. The correlation coefficients and ratios (slopes of the regression lines) between the areas of the two datasets are also annotated in Figure \ref{fig:fig2} with their standard deviations for both the projected and corrected areas. The regression lines were defined as passing through the origin, and their slopes were derived using the method of \citet{1990ApJ...364..104I}, as in many studies, including \citet{2020A&A...640A..78M}. Two regression slopes are possible, depending on the selection of the independent variable: in the case of the MANDAL-NEWSUN combination, either MANDAL or NEWSUN can be the independent variable. If we express the slopes as $b_\mathrm{xy}$ and $b_\mathrm {yx}$ corresponding to the selection of the independent variable, then $b=(b_\mathrm {xy}+1/b_\mathrm {yx})/2$ is adopted as the slope of a regression line.

As presented in Figure \ref{fig:fig2}, the correlation coefficients between MANDAL and NEWSUN/SFTT4 are as high as 0.96--0.97 for the corrected areas and those for the projected areas are still high. These coefficients are close to the best values in the analysis by \citet{2020A&A...640A..78M}. The high correlation coefficients indicate that the area data obtained at Mitaka are of good quality. However, the area ratios are smaller than one, namely, the areas of NEWSUN and SFTT4 are systematically smaller than those of MANDAL.

% Figure 1

% Figure 2

\subsection{Correlation Coefficients and Area Ratios between Various Combinations of the Datasets}

As mentioned before, the correlations between MANDAL and NEWSUN/SFTT4 are high but not perfect and there are systematic differences in the sunspot areas. As mentioned in Section \ref{sec:data}, the observation times of MANDAL were substantially different from those of NEWSUN/SFTT4, and these datasets were processed using a different type of area measurement methods from that of MANDAL. Therefore, we discuss the effect of differences in observation times and area measurement methods on the correlation coefficients and area ratios. For this purpose, we calculated the correlation coefficients and area ratios between SFTT4 and the other datasets listed in Table \ref{tbl:tbl1}. In addition, a special combination of datasets, NEWSUN-MTKDPD, which were processed using different types of methods based on identical data, were also calculated. Because the datasets covered different periods, we used data taken in 2014 to compare the data from the same period. However, the number of MTKDPD samples was much less than those of the other samples. Therefore, we used data from 2012--2015 to compare MTKDPD with NEWSUN or SFTT4.

\begin{figure} 
\centerline{\includegraphics[width=1\textwidth]{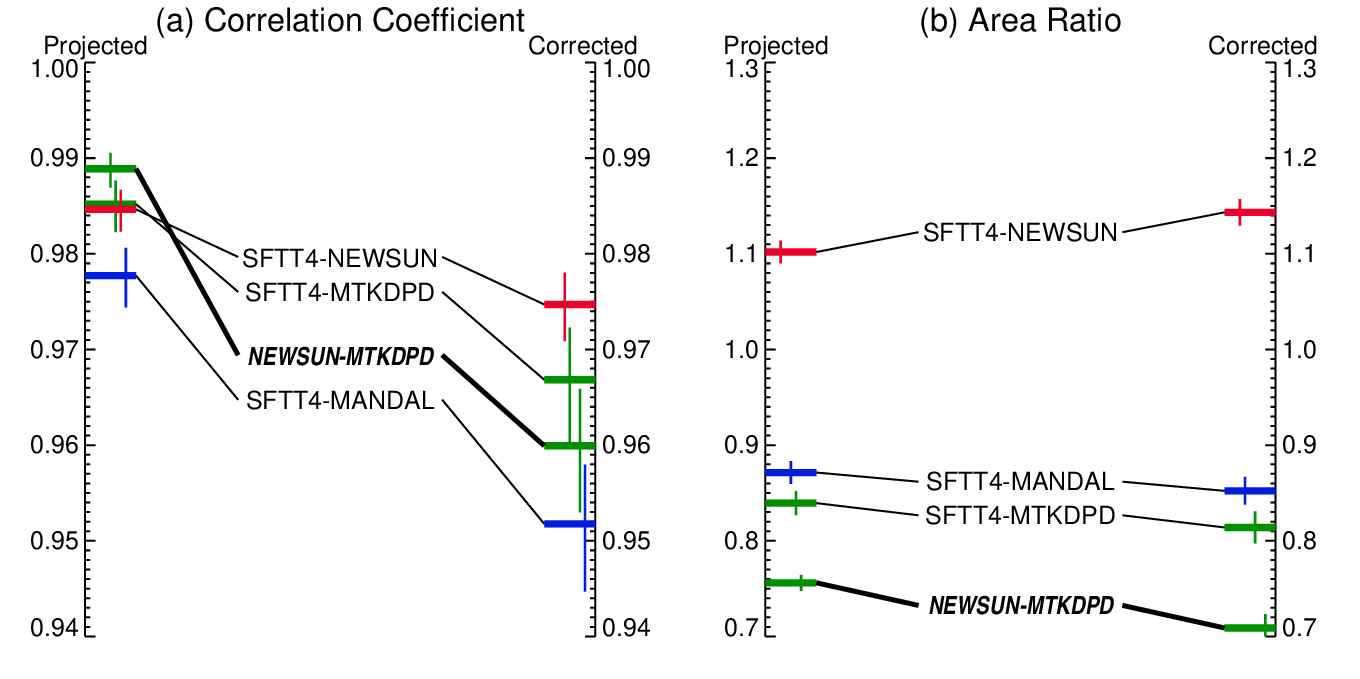}}
\caption{(a) Correlation coefficients and (b) area ratios for the projected and corrected areas between SFTT4 and the other records. The NEWSUN-MTKDPD combination is also presented. The types of data combinations are indicated with three different colors; red for the combination of data taken during the same time period of day processed with similar methods, green for the combinations of data taken during the same time period of day processed with different types of methods, and blue for the combination of data taken during different time periods of day processed with different types of methods. The ranges of $\pm$standard deviations are indicated by vertical bars.}\label{fig:fig3}
\end{figure}

Calculated correlation coefficients and area ratios for projected and corrected areas (with their range of $\pm$standard deviations) are illustrated in Figure \ref{fig:fig3}, where the types of data combinations are indicated with three different colors as explained below. The numbers of samples (common observation days) are shown in parentheses.
\begin{itemize}
\item SFTT4-NEWSUN (199): Comparison of data obtained during the same time period on common observation days and processed with similar automated methods based on thresholding; the red horizontal bars in Figure \ref{fig:fig3}.
\item SFTT4-MTKDPD (123) and NEWSUN-MTKDPD (152): Comparisons of data obtained during the same time period on common observation days and processed using different types (automated or traditional) of methods; green in Figure \ref{fig:fig3}. The combination using identical data, NEWSUN-MTKDPD, is emphasized in this figure.
\item SFTT4-MANDAL (205): Comparison of data obtained during different time periods on common observation days and processed with different types of methods; blue in Figure \ref{fig:fig3}.
\end{itemize}

% Figure 3

Figure \ref{fig:fig3}(a) presents the correlation coefficients. Green or red coefficients for data obtained during the same time period of day are higher than the blue ones for data obtained during different time periods. This suggests that time-induced changes in sunspot shapes reduce the correlation coefficients.

Figure \ref{fig:fig3}(a) shows that the correlation coefficients for the corrected areas are lower than those for the projected areas. Specifically, datasets processed with different types of methods (green and blue) experienced greater reductions in correlation coefficients compared to those processed using the similar automated methods (red). Even when the solar images used were identical (NEWSUN-MTKDPD), this tendency remains the same. This indicates that the difference in the type of area measurement methods---traditional or automated---caused a decrease in the correlation coefficients for the corrected areas. This will be discussed in the following subsection. Nonetheless, the correlation coefficients are high, $>0.95$, even for the corrected areas. 

Figure \ref{fig:fig3}(b) displays the area ratios. The SFTT4 to MANDAL/MTKDPD ratios are comparable in both projected and corrected areas, confirming that datasets processed with traditional methods yield similar areas. Additionally, the ratios of SFTT4 to the other two datasets are less than one. Ratios of NEWSUN to MTKDPD are also below one, indicating that areas calculated by automated thresholding methods are consistently smaller than those obtained through traditional methods. This observation is further discussed in the next subsection.

\subsection{Investigation of the Behavior of Correlation Coefficients and Area Ratios}\label{sec:behaviorccar} 

\subsubsection{Dependence of Group Areas on the Angular Distance from the Disk Center}\label{sec:angulardistance} 

To investigate the cause of the behavior of the correlation coefficients and area ratios shown in Figure \ref{fig:fig3}, we first compared the areas of individual sunspot groups in MTKDPD and NEWSUN. DHO published the area data of individual sunspots and sunspot groups in their DPD catalog; they performed sunspot grouping according to the definitions of active regions by the National Oceanic and Atmospheric Administration (NOAA). We derived the areas of sunspot groups for the data from 2014 to 2015 in the NEWSUN datasets in the same manner as that for MTKDPD. In total, 229 sunspot groups were detected on 34 common observation days in these datasets.

\begin{figure} 
\centerline{\includegraphics[width=1\textwidth]{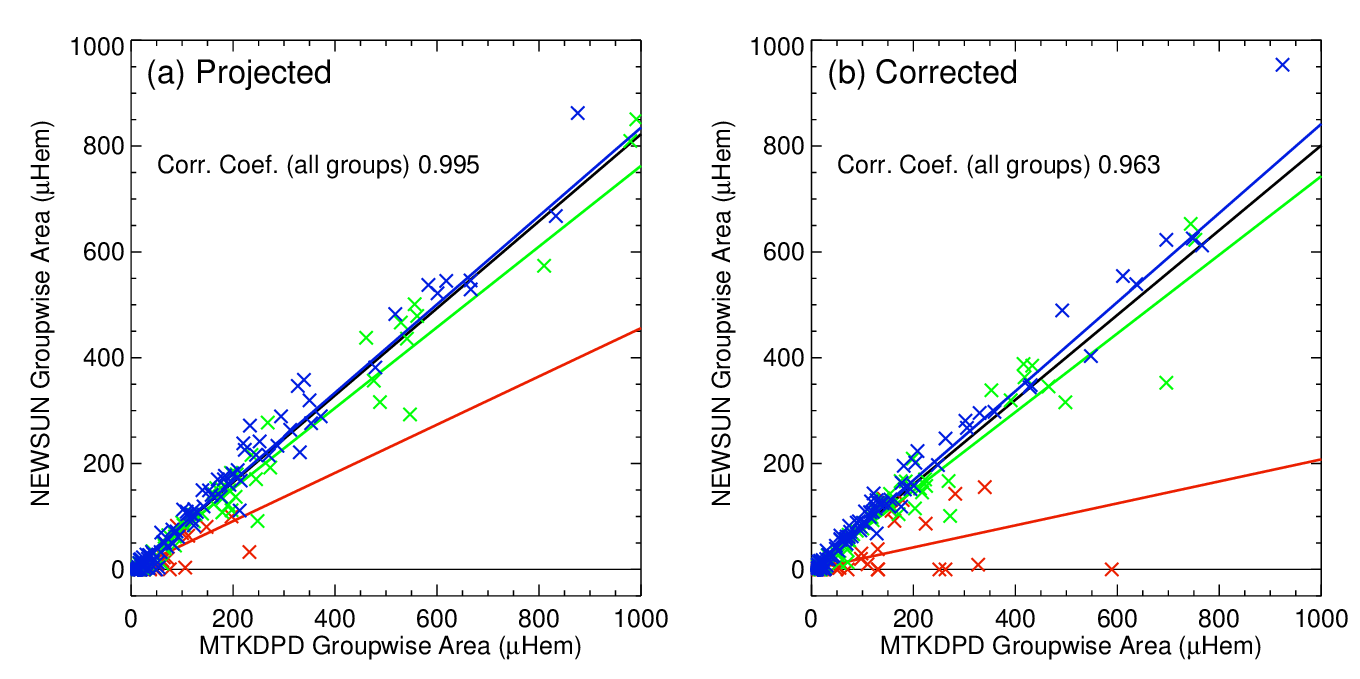}}
\caption{Comparisons of groupwise sunspot-areas between MTKDPD and NEWSUN for (a) projected and (b) corrected cases. To show the behavior of average-sized groups, only the range of 0--1000 $\mu$Hem is shown. The groups presented with cross symbols were color-coded by their angular distance $\theta$ from the disk center; blue for $\theta < 40^\circ$, green for $40^\circ < \theta < 70^\circ$, and red for $70^\circ < \theta $. The regression lines for the points in the three categories were drawn in the corresponding color. The black line in each panel shows the regression line calculated using all the points. The correlation coefficients calculated using all the points were annotated. A thin horizontal line in each panel corresponds to that NEWSUN area = 0. The groups plotted on this line were not detected in NEWSUN.}\label{fig:fig4}
\end{figure}

Figure \ref{fig:fig4} shows scatter diagrams of the groupwise sunspot areas of the MTKDPD-NEWSUN combination for the projected (panel (a)) and corrected (panel (b)) cases. Some exceptionally large groups existed, whose maximum projected and corrected areas reached approximately 4000 and 2000 $\mu$Hem, respectively. However, only data points less than 1000 $\mu$Hem are shown in Figure \ref{fig:fig4} to exhibit the behavior of the average-sized groups. The groups are presented with cross symbols of different colors depending on the angular distance from the disk center $\theta$: blue for $\theta < 40^\circ$, green for $40^\circ < \theta < 70^\circ$, and red for $70^\circ < \theta $. The regression lines for the points in these three categories are shown in the corresponding colors. The thick black lines indicate the regression lines calculated using all the groups (including the large unshown groups). 

The correlation coefficients were as high as 0.995 and 0.963, as shown in the Figures \ref{fig:fig4}(a) and \ref{fig:fig4}(b). The regression lines of the blue and green points (blue and green lines, respectively) were close to the black lines. However, most of the red points ($70^\circ < \theta$) were located below the black lines. In particular, in Figure \ref{fig:fig4}(b), we can see that some red points are on or near the thin horizontal line, which indicates that the area in NEWSUN is zero. This means that groups near the limb tend not to be detected in NEWSUN or are detected as smaller groups than those in MTKDPD. The regression lines of the red points alone (red lines) deviate significantly from the other lines, particularly in Figure \ref{fig:fig4}(b). It is concluded that this larger deviation in Figure \ref{fig:fig4}(b) than that in Figure \ref{fig:fig4}(a) causes the reduction of the correlation coefficient of all the groups from projected areas to corrected areas (0.995 $\rightarrow$ 0.963) because the correlation coefficient of blue and green points alone ($\theta < 70^\circ $) shows only a slight decrease (projected: 0.995 $\rightarrow$ corrected: 0.993). The reduction of the correlation coefficient of all the groups is consistent with that of the daily total areas of NEWSUN-MTKDPD presented in Figure \ref{fig:fig3}(a) (projected: 0.988 $\rightarrow$ corrected: 0.960). For the combination of SFTT4 and MANDAL, we also confirmed that the exclusion of the groups of $\theta > 70^\circ $ improved the correlation.

% Figure 4

\subsubsection{Detailed Comparison of Group Areas between Datasets}\label{sec:grouparea} 

\begin{figure} 
\centerline{\includegraphics[width=1\textwidth]{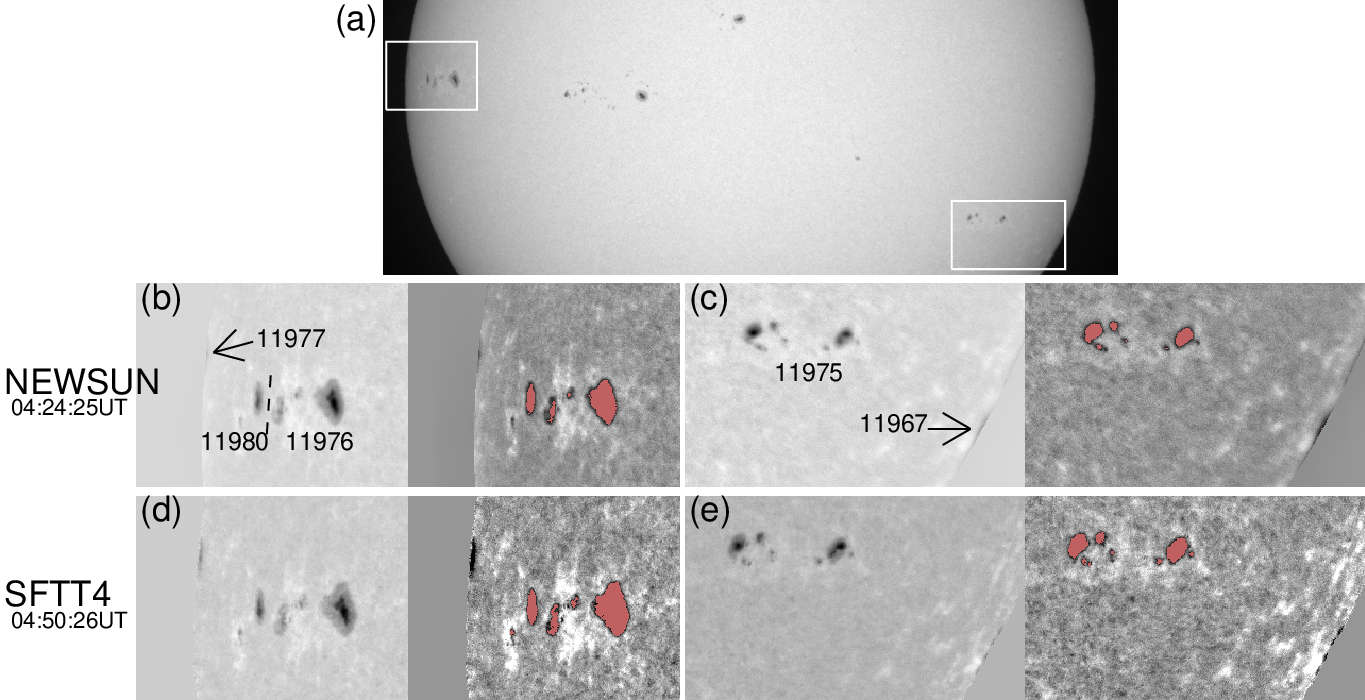}}
\caption{Spot-detection results of NEWSUN and SFTT4 on 2014 February 10. Close-up views of regions near the limb, whose positions are indicated in panel (a), are shown in panels (b)--(e). Panels (b)(c) and (d)(e) show the NEWSUN and SFTT4 images, respectively. In all images, the quiet disk component was removed. In the right-side images in panels (b)--(e), the regions distinguished as sunspots are displayed in red. NOAA active region numbers of the groups are indicated in panels (b) and (c).}\label{fig:fig5}
\end{figure}

To demonstrate how the spots near the limb were treated by automated methods, we presented the spot-detection results of NEWSUN and SFTT4 on 2014 February 10 in Figure \ref{fig:fig5} as typical examples. This day was included in MTKDPD; the DPD catalog used the NEWSUN image on this day to measure sunspot areas. Close-up views of the regions near the limb, whose positions are indicated in Figure \ref{fig:fig5}(a), are shown in Figures \ref{fig:fig5}(b)--(e). NOAA active region numbers are indicated in panels (b) and (c). Panels (b)(c) and (d)(e) show the NEWSUN and SFTT4 images, respectively. In all images, the quiet disk component was removed. The right-side images in panels (b)--(e) are shown in higher contrast, and the regions distinguished as sunspots are displayed in red. NOAA 11977 and 11967 were located at the east limb and west limb, respectively, and sunspots in these regions were barely visible. Although these sunspots were not identified in NEWSUN (Figures \ref{fig:fig5}(b) and 5(c)), they were identified in MTKDPD, which used the same image. No sunspots were identified in SFTT4, either. The dark features in NOAA 11977 and 11967 seen in Figures \ref{fig:fig5}(b)--(e) were judged as false spots by the automated methods because they could not be distinguished from a part of the ragged limb produced by the seeing effect.

% Figure 5

\begin{figure} 
\centerline{\includegraphics[width=1\textwidth]{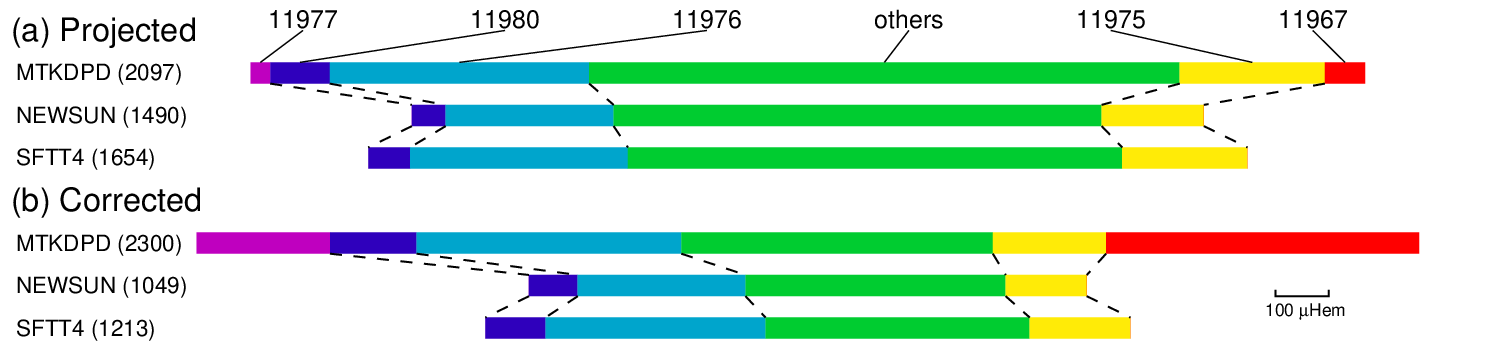}}
\caption{Bar charts of (a) projected and (b) corrected areas of the individual groups on 2014 February 10 measured in MTKDPD and the two datasets shown in Figure \ref{fig:fig5}. Each group is shown in a different color and each bar length denotes a group area. Besides the regions near the east limb and west limb seen in Figure \ref{fig:fig5}, other active regions absent in Figure \ref{fig:fig5}(b)--(e) are grouped as ``others''. Daily total areas ($\mu$Hem) are indicated in parentheses.}\label{fig:fig6}
\end{figure}

Figure \ref{fig:fig6} displays bar charts of the areas of sunspot groups presented in Figure \ref{fig:fig5}, extracted from MTKDPD, NEWSUN, and SFTT4. The horizontal bar lengths denote the areas, with each group color-coded. Some active regions in Figure \ref{fig:fig5}(a) are absent in Figures \ref{fig:fig5}(b)--(e), and are grouped as ``others'' in Figure \ref{fig:fig6}. In Figure \ref{fig:fig6}(a), the projected area shows that NOAA 11977 and 11967 at the limb constituted only 5 \% of MTKDPD's total area. NEWSUN and SFTT4's exclusion of these regions does not notably affect the total projected areas. However, the area in MTKDPD for these regions swell up to 37 \% after the foreshortening correction (panel (b)). 

This result indicates that detecting spots very close to the limb is challenging not only for NEWSUN but also for SFTT4, despite improvements in the automated detection for SFTT4. This detection difficulty causes discrepancies in corrected areas between automated and traditional methods, contributing to lowering of the correlation coefficients from projected to corrected daily total areas for datasets processed by different types of methods, as shown in Figure \ref{fig:fig3}(a) (green or blue horizontal bars).

% Figure 6

\subsubsection{Cause of Systematic Difference in Sunspot Areas}\label{sec:sysdif} 

Notably, the areas derived using automated methods at Mitaka are consistently smaller than those obtained through traditional methods. As illustrated in Figure \ref{fig:fig3}(b), the areas for SFTT4 and NEWSUN are smaller compared to those derived by traditional methods, MANDAL and MTKDPD. While Figure \ref{fig:fig3} presents data from a limited period (primarily in 2014), this trend is also evident in Figure \ref{fig:fig2}, which covers extended periods for NEWSUN and SFTT4. 

The data shown in Figures \ref{fig:fig5} and \ref{fig:fig6} reveal a similar pattern. Comparing the total areas of groups (excluding those undetected by NEWSUN/SFTT4) between MTKDPD and NEWSUN/SFTT4 indicates that NEWSUN and SFTT4 have total projected areas of 75 \% and 83 \% of MTKDPD, and 72 \% and 83 \% of the corrected areas, respectively. These ratios closely match those in Figure \ref{fig:fig2}, with 75 \% and 85 \% for the projected area, and 70 \% and 83 \% for the corrected area. Thus, the spots identified by automated methods, shown as red patches in Figure \ref{fig:fig5}, are considered typical examples presenting smaller areas than those identified by traditional methods.

Examining the right-side images in Figures \ref{fig:fig5}(b)--(e) reveals red patches with dark borders. This indicates that the automated methods at Mitaka do not include the outermost edges of the spots in the derived spot areas. 

In automated thresholding methods, a specific brightness threshold is set, distinguishing regions below this threshold as sunspots. Typically, the threshold is significantly lower than the average brightness of quiet disk to avoid misidentifying the darker parts of granules as sunspots. Consequently, the penumbral boundaries identified by automated thresholding fall inside those defined by traditional methods, which rely on human observation or emulation thereof. Thus, the areas calculated by automated methods at Mitaka are consistently smaller than those obtained through traditional techniques. 

Systematic differences between sunspot areas determined by automated and traditional methods can be calibrated using a correction factor to align with reference data. High correlation coefficients confirm the validity of this approach. Consequently, despite these differences, automated method data from Mitaka can be included in the sunspot-area time series.

\section{Summary and Discussion} \label{sec:summary}

The daily sunspot-area data obtained at the Mitaka campus of NAOJ using the automated full-disk sunspot-detection methods based on thresholding, namely NEWSUN and SFTT4, show high correlations with the data in the reference catalog by \citet{2020A&A...640A..78M}, which is a composite of various calibrated sunspot area records. Although the correlation coefficients are high, they are limited by factors other than random errors and the differences in instruments and seeing. First, the difference in observation times in the day reduces the correlation to some extent. Because the sunspot area varies even within a day, the daily area values may differ depending on the observation time. This is an inevitable uncertainty in the daily sunspot area data. Second, in the automated detection methods at Mitaka, locating sunspots that are very close to the limb is challenging, because of the blurring due to the seeing effect. By contrast, such sunspots can be identified using the traditional sunspot-area measurement methods. This causes a discrepancy between the results of the automated and traditional methods, particularly in areas corrected for the foreshortening effect, and reduces the correlation. 

It should be noted that the areas of sunspots very close to the limb detected by the traditional methods are inflated at the time of conversion to the corrected areas, as shown in Figure \ref{fig:fig6}. This implies that a small error or change in the projected areas of such sunspots may produce large differences in the corrected areas. The appearance of spots close to the limb changes rapidly with the solar rotation. Therefore, a large scatter may be observed in the corrected areas captured at different times of the day, particularly when measured using traditional methods.

Meanwhile, the areas derived by the automated methods are systematically smaller than those by the traditional methods. The outline of penumbrae is determined by a threshold for sunspot detection in the automated methods, and it is generally inside of the visually determined outline, which is used in the traditional methods; this is the reason of the systematic discrepancy. 

Systematic discrepancies in sunspot areas between records obtained from different observatories have been a long-standing problem. In particular, the RGO and the SOON data reportedly show a large discrepancy; the SOON areas should be increased by a factor of 1.4 to match the RGO data \citep{1997SoPh..173..427F, 2002SoPh..211..357H, 2009JGRA..114.7104B}. Initially, this discrepancy was presumed to be caused by the exclusion of small spots \citep{2014SoPh..289.1517F}, but it was pointed out that the SOON data actually include small spots \citep{2017MNRAS.465.1259G, 2018SoPh..293..142B}. Some reanalyses of the SOON data presented larger areas than the original measurements \citep{2018SoPh..293..138G, 2020MNRAS.497.1110M}. The reason for this discrepancy between RGO and SOON remains unclear. However, the differences between the areas derived using the automated methods at Mitaka and those derived using the traditional methods are primarily ascribed to the threshold. Because the cause of the systematic difference is clear, and the correlation is high, a correction factor can be applied to calibrate the areas, determined using the automated methods, to match those in the catalog reported by \citet{2020A&A...640A..78M}.

NEWSUN and SFTT4, which are processed using the automated methods, show the common tendencies mentioned before. However, a comprehensive examination of the correlation coefficients indicates the existence of differences between SFTT4 and NEWSUN. In the comparison of the corrected area with MANDAL, SFTT4 shows a slightly higher correlation than does NEWSUN (Figure \ref{fig:fig2}(b)). The correlation between SFTT4 and MTKDPD for the corrected area is higher than that between NEWSUN and MTKDPD (Figure \ref{fig:fig3}(a)) because of the strong capability of SFTT4 to detect spots close to the limb, whereas the NEWSUN-MTKDPD correlation is higher than the SFTT4-MTKDPD correlation for the projected areas. As discussed before, the systematic differences in the areas can be calibrated. Therefore, SFTT4 can be considered a useful dataset for extending the sunspot area catalog.

Traditional methods provide similar areas, irrespective of the instruments used to capture the images. Therefore, area measurements using traditional methods should be continued with the aid of several observatories. However, in reality, this is challenging because the process of deriving spot areas is both time- and labor-intensive. Therefore, automated methods based on thresholding, which can be carried out in less time and effort and exhibit a high correlation with the reference data, are also necessary for the continuation of sunspot-area measurement. Of course, it is important to monitor the time variation of the correlation and area ratios between automated and traditional methods. The overlap between SFTT4 and MANDAL is rather short, and therefore, it is crucial to compare the Kislovodsk or SOON data for further investigation.

In addition, the results of automated detection show a brightness deficit owing to sunspots. Such data can be used to verify the PSI, which are used to reconstruct the solar irradiance variation before the digital observation era. The brightness deficits are presented in Appendix A.

%% Figure 
%
% \begin{figure} 
% \centerline{\includegraphics[width=0.5\textwidth,clip=]{<fig.eps>}}
% \caption{}%\label{fig:?}
% \end{figure}

%% Table
%
% \begin{table}
% \caption{}%\label{tbl:?}
% \begin{tabular}{}     
% \hline
% \multicolumn{2}{c}{<>}
% <data>
% \hline
% \end{tabular}
% \end{table}

%%%%%%%%%%%%%%%%%%%%%%%%%%%%%%%%%%%%%%%%%%%%%%%%%%%%%%%%%%%%%%%%%%%%%%%%%%%
%% Appendix
%
\appendix   

\section{Brightness Deficit Due to Sunspots and the PSI} \label{sec:appendix}

The PSI \citep{1982SoPh...76..211H} is an index representing brightness deficit due to sunspots, and it is calculated from sunspot areas based on an empirical relation between the sunspot area and the brightness deficit. The catalog by \citet{2020A&A...640A..78M} listed the PSI along with the sunspot areas.

\begin{figure} 
\centerline{\includegraphics[width=1\textwidth]{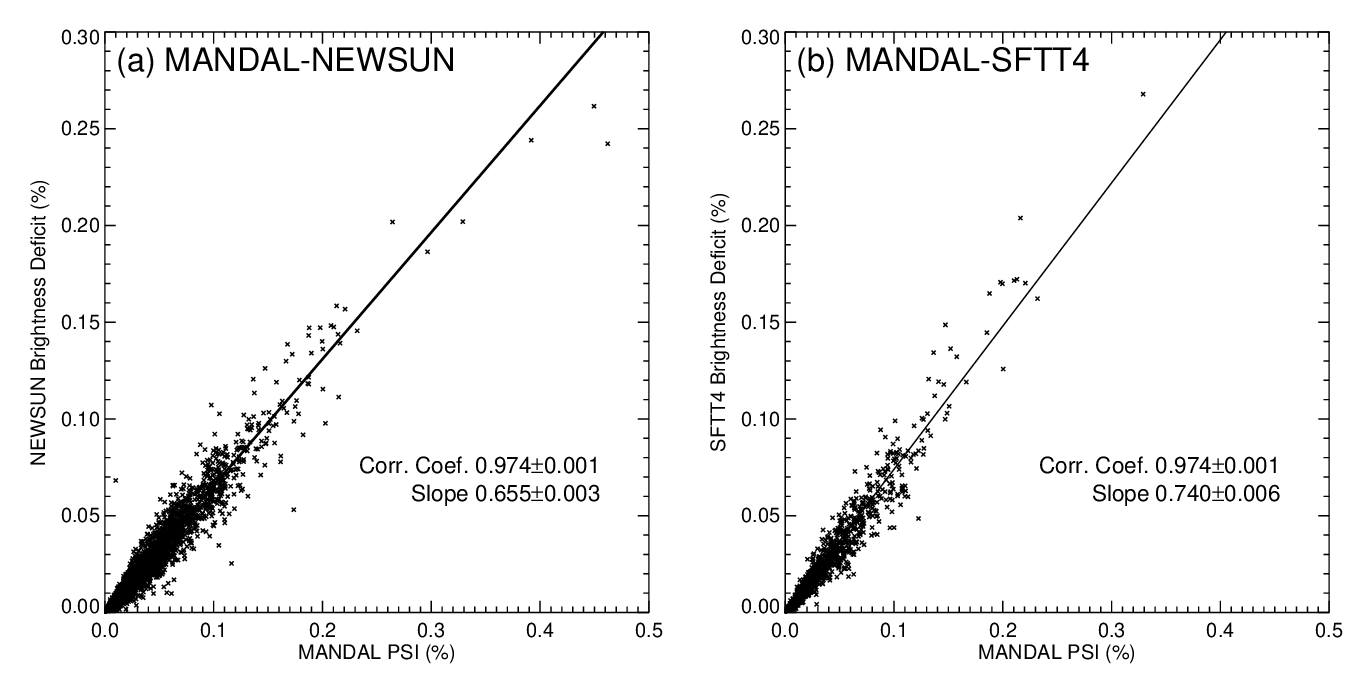}}
\caption{Scatter diagrams of the PSI derived by \citet{2020A&A...640A..78M} and the brightness deficit measured in (a) NEWSUN and (b) SFTT4. The correlation coefficients and slopes of the regression lines were annotated in the Figure.}\label{fig:fig7}
\end{figure}

Automated sunspot detection methods based on digital images employed at Mitaka provide the amount of directly measured brightness deficits as well as sunspot areas. Here, we compare the PSI and brightness deficit. Figure \ref{fig:fig7} shows scatter diagrams of the PSI in MANDAL and the brightness deficit measured in NEWSUN and SFTT4. The correlation coefficients and slopes of the regression lines were annotated. The PSI is linearly related to the brightness deficit, and the correlation coefficients in both cases are as high as 0.974. However, the brightness deficit is systematically smaller than that of the PSI, as is the sunspot area.

We did not consider the effects of scattered light, which may have reduced the apparent deficit. Furthermore, to discuss irradiance variation, the excess brightness due to faculae should also be studied. Therefore, our comparison is preliminary; however, it helps to discuss the validity of the PSI as a proxy for the brightness deficit due to sunspots. \citet{2013SSRv..176..237F} argued that the expression for the PSI proposed by \citet{1996ApJ...461..478S} is more appropriate. Further comprehensive studies on the PSI and the brightness deficits are imperative.

% Figure 7

%%%%%%%%%%%%%%%%%%%%%%%%%%%%%%%%%%%%%%%%%%%%%%%%%%%%%%%%%%%%%%%%%%%%%%%%%%%
% Acknowledgements
\begin{acks}
We thank DHO, who made a great endeavor to produce the DPD catalog and make it available online.
\end{acks}

%% Available additional data environments:
%% required: authorcontribution, fundinginformation, dataavailability
%% optional: materialsavailability, codeavailability
\begin{authorcontribution}
Conceptualization, analysis, and investigation were performed by Y. H. He wrote the manuscript as well.
\end{authorcontribution}
\begin{fundinginformation}
No funding was received for conducting this study.
\end{fundinginformation}
\begin{dataavailability}
The authors declare that the data used this study are available in a web page of the Solar Science Observatory of the National Astronomical Observatory of Japan (\burl{https://solarwww.mtk.nao.ac.jp/en/database.html}).
\end{dataavailability}
\begin{ethics}
\begin{conflict}
The authors declare no competing interests.
\end{conflict}
\end{ethics}

%%% %%%%%%%%%%%%%%%%%%%%%%%%%%%%%%%%%%%%%%%%%%%%%%%%%%%%%%%%%%%
%% Bibliography
%
% Without BibTeX 

\end{document}